\documentclass[]{iopjournal}

\usepackage{amsmath,amssymb,amsfonts}
\usepackage{bm}
\usepackage{mathrsfs}
\usepackage{dsfont}
\usepackage{graphicx}
\usepackage{hyperref} 
\hypersetup{colorlinks=true,linkcolor=blue,citecolor=blue,urlcolor=blue}
\usepackage{color,soul}

\newcommand{\be}{\begin{equation}}
\newcommand{\ee}{\end{equation}}
\newcommand{\bea}{\begin{eqnarray}}
\newcommand{\eea}{\end{eqnarray}}

\begin{document}

\articletype{Paper} 

\title{Noise-limited secret key agreement with twin optical physically unclonable functions}

\author{Georgios M. Nikolopoulos$^{1,2,*}$\orcid{0000-0002-3937-2771}}

\affil{$^1$Institute of Electronic Structure \& Laser (IESL), FORTH, GR-70013 Heraklion, Greece}

\affil{$^2$Center for Quantum Science \& Technologies (FORTH-QuTech), GR-70013 Heraklion, Greece}

\affil{$^*$Author to whom any correspondence should be addressed.}

\email{nikolg@iesl.forth.gr}


\date{\today}

\begin{abstract}
\noindent
We investigate the use of twin optical fingerprints derived from correlated physical unclonable functions (PUFs), as a hardware-based platform for cryptographic key generation and distribution. Each fingerprint is associated with a random, yet reproducible speckle pattern, generated when coherent light is scattered by a disordered optical structure. We consider a pair of correlated optical PUFs, and study the conditions under which two honest parties can establish a common secret key, despite fabrication-induced variability and environmental noise. An explicit information-theoretic key-agreement protocol is developed, incorporating secure sketches, error reconciliation, and privacy amplification. We quantify information leakage due to public helper data, and derive lower bounds on the length of the final secret key. The analysis identifies the noise regimes in which secure key agreement is feasible, and examines the performance of both practical and near-capacity reconciliation schemes. Finally, we discuss how twin optical PUFs could be integrated into quantum key distribution (QKD) networks,  as a mechanism for establishing an initial pre-shared secret key between two honest users, without relying on computational assumptions or trusted third parties.
\end{abstract}

\section{Introduction}
\label{sec1}

Physical Unclonable Functions (PUFs) are hardware security primitives that exploit the inherent randomness introduced during the manufacturing process of physical systems \cite{PUF-review1,PUF-review2,PUF-review3}. 
This randomness is to some extent uncontrollable even by the manufacturer, and gives rise to unique and irreproducible physical characteristics that can be harnessed to generate device-specific cryptographic keys or fingerprints. 
Unlike traditional digital keys, which can be extracted or duplicated, PUFs offer intrinsic resistance to cloning and tampering, making them highly attractive for a range of critical security applications  \cite{PUF-review4,PUF-review5,PUF-review6,PUF-review7}.

Among the various PUF implementations, optical PUFs (O-PUFs) represent a compelling subclass due to their simplicity, passive nature, and high entropy. 
These systems typically rely on the random scattering of light through disordered media, such as polymers doped with  randomly distributed  air bubbles or nanoparticles \cite{pappu2002,Ruh-etal,lio2022,PUF-review8,Wang21}, 
layers of white paint, nanostructured or microstructured surfaces with random features, fused glass materials with random defects \cite{Iesl16}, 
disordered or multicore optical fibers \cite{Mes18,Akr18}, etc. 
Recent advances in optical PUF technologies include plasmonic metasurface-based architectures, structural-color-based unclonable devices, and reconfigurable  O-PUF platforms that enable dynamic generation of multiple optical fingerprints \cite{Kim2025,Lin2026,Kwak2026}.
In O-PUFs, the disordered medium plays the role of a random token, and in response to a light-based challenge, 
it produces a complex random speckle. The speckle is a “fingerprint” that characterizes the O-PUF, and it is technologically hard to duplicate, without access to the token.  

Applications of O-PUFs span both classical and emerging security domains. In authentication, O-PUFs can be embedded into products to verify their origin and integrity \cite{buch05}, while they can be also used in smart cards for entity authentication purposes \cite{pappu2001,pappu2002,Iesl16,Mes18,Akr18}. 
In secure key storage, they offer a non-digital method to store and retrieve cryptographic keys on demand, avoiding risks associated with stored secrets \cite{Horstmayer13}. 
Moreover, O-PUFs are increasingly explored in the design of quantum-safe protocols for entity authentication  \cite{Goorden14,NikDiaSciRep17,Nik21}, encryption \cite{uppu19, Rates23}, commitments \cite{Nik19}, etc, which offer high level of security against both classical and quantum adversaries  \cite{PUF-review5,Gia20}.  
Most of these applications rely on challenge–response mechanisms involving a single PUF instance, where security is derived from the difficulty of predicting or reproducing the response of the device. The establishment of a common secret key between two users is considerably more challenging, as it requires the generation of correlated responses at different locations. Existing approaches typically rely on trusted enrollment procedures, stored challenge–response information  \cite{NikFis24,Konteli26}, on the transfer of a PUF between users under trusted conditions \cite{Brz11}, or the transmission of optical challenges between users.

The twin-PUF paradigm offers an alternative route to secret-key establishment, in which the required correlation between the two parties is embedded directly into their physical devices during fabrication. In the present work, we consider twin O-PUFs, that is, pairs of PUF instances that are intentionally fabricated under tightly controlled and identical process conditions, with the goal of minimizing fabrication-induced variability. 
Although these PUFs are not physically identical, they exhibit highly correlated challenge–response behavior due to the shared fabrication environment. 
The fabrication of such twin PUFs has been experimentally demonstrated using high-resolution nanoscale 3D printing \cite{Mar22}. Importantly, the concept of twin PUFs does not contradict the principle of unclonability, as it differs fundamentally from adversarial cloning. 
Cloning typically refers to an unauthorized attempt to replicate a given PUF instance—either physically or via modeling. 
In contrast, twin PUFs are manufactured by a trusted entity  under the same conditions, and are not exact copies, but statistically similar devices.

The experimental demonstration of twin  O-PUFs reported in Ref. \cite{Mar22} established the feasibility of fabricating physically distinct optical structures with  highly correlated responses. This capability opens the possibility of using twin PUFs as a source of shared physical randomness for cryptographic applications, including secret-key generation and secure communication. However, while Ref. \cite{Mar22} discusses such potential applications, it does not address the information-theoretic conditions under which two users can reliably extract a common secret key from correlated PUF instances, nor does it quantify the impact of noise, reconciliation overhead, helper-data leakage, or privacy amplification on the achievable secret-key length.
In this work, we address these questions by investigating theoretically the conditions under which twin O-PUFs can be employed for the generation and distribution of a secret key between two honest parties. Our analysis accounts for fabrication-induced variability, environmental noise, and practical considerations related to error reconciliation and public communication between users. In particular, we explicitly account for information leakage through publicly exchanged helper data and derive quantitative lower bounds on the length of the final secret key using information-theoretic tools.
 In contrast to Refs. \cite{NikFis24,Konteli26,Brz11}, the protocol proposed here derives a shared secret directly from the correlated responses of two distinct twin PUFs, without requiring stored challenge–response databases or physical transfer of the devices.
Finally, we explore the integration of twin O-PUFs into quantum key distribution (QKD) systems as a practical method for establishing an initial pre-shared key. This approach addresses a central challenge in QKD, namely the need for initial key establishment, without relying on computational assumptions.

The fabrication and use of twin PUFs has also been  experimentally demonstrated by Zhong et al. \cite{Zhong22}, who introduced carbon-nanotube–based twin PUFs exhibiting high correlations. In that work, secure communication was demonstrated by directly encrypting plaintext using raw PUF responses and mitigating bit errors through fault-tolerant repetition and majority voting. The focus of Ref. \cite{Zhong22} is primarily on device fabrication, reliability, and error suppression at the hardware level. 
The present work is focused on optical PUFs and addresses a complementary but distinct problem. Rather than relying on direct encryption with noisy PUF outputs, we formulate an explicit, rather general, cryptographic key-agreement protocol based on twin PUFs, incorporating secure sketches, error reconciliation, and privacy amplification. This allows us to quantify information leakage due to public helper data, and to derive information-theoretic lower bounds on the final secret-key length using min-entropy arguments. 
Moreover, we introduce a general stochastic model for fabrication-induced and environmental noise that is independent of a specific PUF platform, and we analyze the feasibility of key agreement as a function of an experimentally accessible quantity, namely the disagreement probability (bit-error rate) between the binary keys extracted from the PUF responses. Although the present work is motivated by optical twin PUFs, the resulting information-theoretic framework can, in principle, be extended to other correlated PUF platforms that generate sufficiently similar binary keys, including the CNT-based twin PUFs of Ref. \cite{Zhong22}. To our knowledge, this is the first work that combines a physical model of correlated twin PUFs with an explicit information-theoretic key-agreement protocol incorporating secure sketches, error reconciliation, privacy amplification, and quantitative secret-key rate analysis under realistic noise assumptions.

\section{Methods}
\label{sec2}

From a mathematical point of view, an O-PUF is a  function that refers to the response of a disordered physical object ${\cal T}$ (to be referred to hereafter as the token), 
when light is scattered from it. 
In the absence of nonlinearities, and for tokens with sufficiently small spectral correlation bandwidth, 
the scattering of monochromatic laser light from the token is a linear process involving a large number of transverse spatial modes. 
At the output, a characteristic speckle pattern emerges due to the interference of the many optical paths that contribute to each spatial mode.
In the diffusive limit  the electric field at the $m$th output mode is given by \cite{pappu2001,Goodman1,Skoric2008} 
\begin{subequations}
\label{speckle_eqs}
\begin{equation}
E_m = \sum_{j=1}^{\cal N} T_{m,j} E_j^{\rm (in)}
\label{Em:eq}
\end{equation}
where $T_{m,j}$ is the element of the transmission matrix connecting the $m$th output mode, to the $j$th input mode, for the wavelength of the input light. 
For a given wavelength, the 
elements of the transmission matrix $\{T_{m,j}\} $ depend strongly on the realization of the disorder in the token, and can be treated as independent complex Gaussian random variables, 
with zero mean and variance given by $\sigma^2 = l/(L{\cal N})$, where ${\cal N}$ is the number of input modes,  $L$ is the thickness of the token and  $l \ll L$ is the mean free path.

Assuming uniform illumination, the electric field at the $j$th input mode can be taken as 
\begin{equation}
E_j^{\rm (in)} = {\cal E} e^{{\rm i} \phi_j}/\sqrt{\cal N}, 
\end{equation}
where the phase $\phi_j$ can be controlled by means of a phase-only spatial light modulator (SLM). For an $N-$bit SLM, we have 
\begin{equation}
\phi_j = \frac{2\pi c_j}{2^N} , 
\end{equation}
where the integer  $c_j\in [0, 2^N)$. The phase mask ${\bm \Phi} :=\{\phi_1,\phi_2,\ldots,\phi_{\cal N}\}$ essentially determines the wavefront of the input light and defines the optical challenge in our model. 
\end{subequations}

An O-PUF operates as a pseudo-random number generator, in the sense that the raw noisy  speckle that is produced as a response to an input stimulus can be converted to a stable numerical key after its discretization (e.g.,  using Gabor filters) and using  a fuzzy extractor. The fuzzy exrtactor typically involves two separate processes namely, 
reconciliation and hashing.  The former aims at a reconciled key which is not affected by environmental  variations and ageing of the token, whereas the latter compresses the reconciled key further, so that the final key is nearly uniformly distributed.   
Given that throughout this work we deal with optical PUFs only, from now on we simply write PUF(s) instead of O-PUF(s), for the sake of brevity. 

For our purposes, we need to model variations in the response of twin PUFs, resulting from both fabrication-induced randomness and environmental fluctuations. To this end, we adopt a stochastic modeling, 
pertaining to a fixed reference transmission matrix $ {\bf T}^{(0)} $. 
Each element of ${\bf T}^{(0)} $ is drawn independently from a circular complex Gaussian distribution with zero mean and variance $\sigma^2$ i.e., $    T^{(0)}_{m,j} \sim \rm{CN}(0, \sigma^2)$.
$ {\bf T}^{(0)}$ is used to generate the individual matrices of the twins, \({\bf T}^{\rm (A)} \) and \({\bf T}^{\rm (B)} \), by adding independent complex Gaussian perturbations
 ${\bf T}^{(u)}= {\bf T}^{(0)} + {\bf W}^{(u)}$, 
where ${\bf W}^{(u)}$  is a perturbation matrix with elements $W_{m,j}^{(u)} \sim \rm{CN}(0, \sigma_{\text{F}}^2)$, and $ \sigma_{\text{F}}<\sigma$.
By varying the standard deviation $\sigma_{\text{F}}$, we can simulate different levels of fabrication variability, and multiple twin pairs are generated for each level.
Using $ {\bf T}^{(0)} $ as a neutral reference ensures that both twins are statistically equivalent and inherit equal, independent deviations from the same underlying physical system. 
In practical fabrication scenarios, neither twin serves as the ``original''; instead, both devices are subject to independent random imperfections during a common manufacturing process. 
Hence, our model reflects the physical symmetry of twin PUF fabrication, where devices are co-fabricated under identical nominal conditions, rather than cloned from one another.
By treating $ {\bf T}^{(0)} $ as a conceptual blueprint and applying symmetric noise to generate $ {\bf T}^{\rm (A)} $ and ${\bf T}^{\rm (B)} $, 
the model guarantees that the deviations are independent and identically distributed, while avoiding bias in pairwise comparisons.
To simulate disturbances associated with thermal or mechanical fluctuations, etc, environmental noise is introduced during each realization of the measurement process :
$ \widetilde{\bf T}^{(u)} = {\bf T}^{(u)}+{\bf V}^{(u)}$, 
where ${\bf V}^{(u)}$  is a perturbation matrix due to environmental noise,  with elements $V_{m,j}^{(u)}\sim \rm{CN}(0, \sigma_{\text{E}}^2)$, 
and standard deviation $ \sigma_{\text{E}}<\sigma$. Multiple realizations (trials) capture variability due to dynamic conditions. 

For each transmission matrix $ \widetilde{\bf T}^{(u)} $, the output speckle  is computed using Eqs. (\ref{speckle_eqs}) with $T_{m,j} = \widetilde{T}_{m,j}^{(u)} $, and the intensity for the $m$th output mode is 
$I_m^{(u)} = |E_m^{(u)}|^2$. 
The speckle pattern is processed with a set of Gabor filters at different frequencies and orientations to extract robust spatial features. Each filter response is binarized independently by median thresholding, and the resulting binary maps are concatenated to form a reproducible numerical key\cite{pappu2001,Mes18,Akr18,Skoric2008,Sha12}.

In closing, we would like to point out that the parameters $\sigma_{\rm F}$ and $\sigma_{\rm E}$ should be interpreted as effective phenomenological measures of fabrication-induced variability and environmental perturbations, respectively. They are not intended to represent directly measurable fabrication tolerances or environmental quantities, such as surface roughness, refractive-index fluctuations, temperature variations, or mechanical vibrations. Rather, they quantify the aggregate effect of all such perturbations on the transmission matrix of the PUF. In practice, these parameters are generally not measured directly. 
Rather, their effect is inferred through observable quantities, such as the similarity between optical responses (e.g., Pearson correlation coefficients of the speckle patterns) or, after feature extraction and binarization, through the normalized Hamming distance between the corresponding binary keys. These observable quantities therefore provide the natural link between the theoretical model and experimental implementations. Consequently, the thresholds identified in the present work should be interpreted as bounds on the effective variability experienced by the system, rather than as direct requirements on any individual fabrication or environmental parameter. The platform-independent nature of the model allows the same framework to be applied across a broad range of PUF technologies, optical and non-optical. Establishing a quantitative mapping between $(\sigma_{\rm F}, \sigma_{\rm E})$, and specific fabrication or environmental parameters would require a dedicated experimental characterization of a particular platform under consideration, and lies beyond the scope of the present work.

\begin{figure*}[!t]
\centering\includegraphics[width=5in]{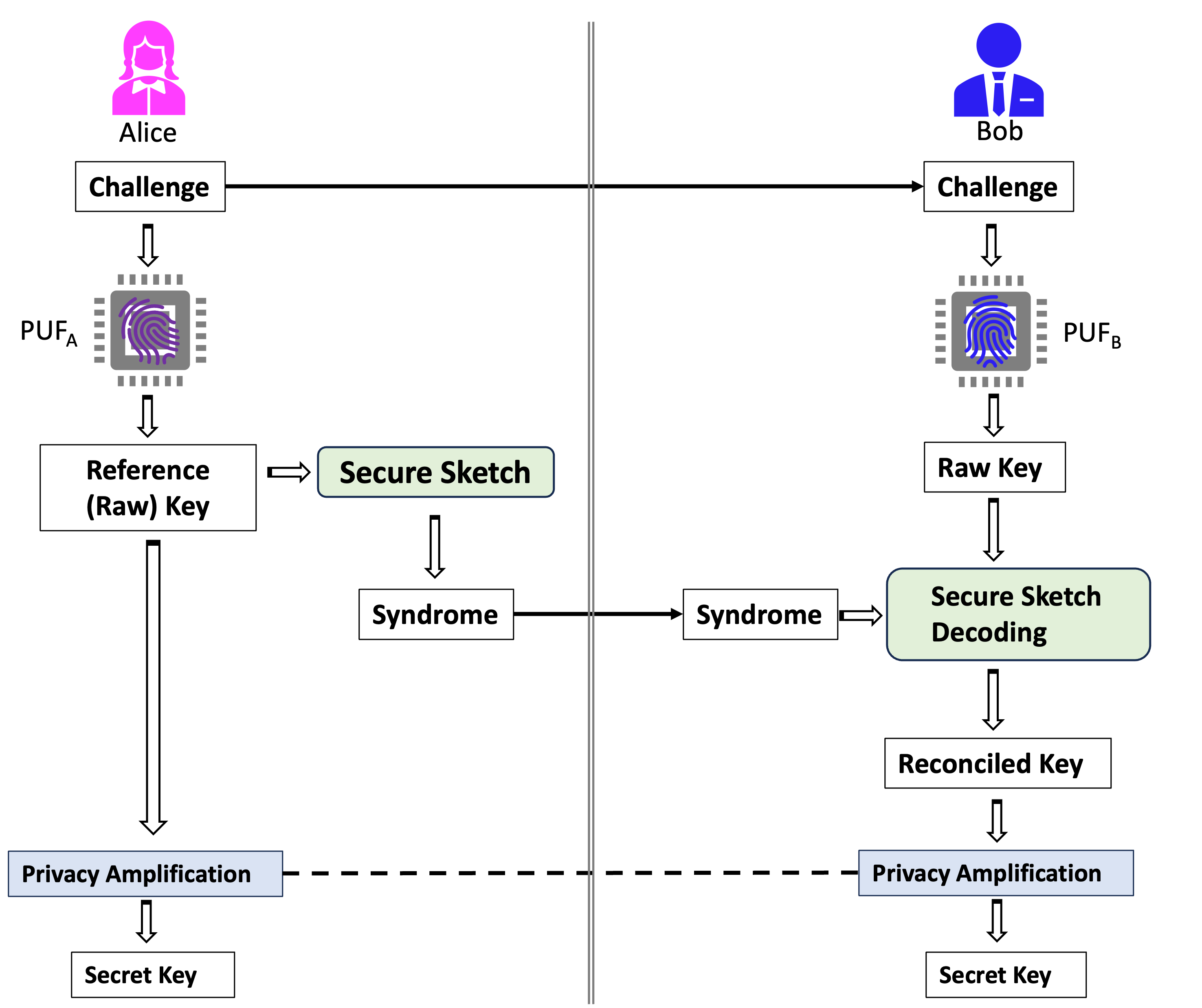}
\caption{Schematic representation of the main steps in the key agreement with  twin PUFs.}
\label{fig_1}
\end{figure*}

\section{Results}
\vspace*{0.2cm}

In this section, we discuss the main results of our work. In particular, we present a protocol for secret-key agreement that exploits correlations between keys produced by twin PUFs. We evaluate its performance in both low- and high-noise regimes, incorporating insights from related experimental studies.

\subsection{Protocol for key agreement with twin PUFs}
\label{sec3}

Our twin-PUF key agreement protocol is schematically illustrated in Fig.~\ref{fig_1}. 
Consider two honest parties, Alice and Bob, possessing the twin PUFs: PUF$_{\rm A}$ and PUF$_{\rm B}$, respectively. 
They simultaneously interrogate their twin tokens ${\cal T}_{\rm A}$ and ${\cal T}_{\rm B}$ with a common challenge input, producing the corresponding optical speckle patterns (not shown in the figure). 
These speckles are processed (e.g., using Gabor filters or other hashing techniques) to yield binary raw keys of length $n$ \cite{pappu2002,Mes18,Akr18,Skoric2008,Sha12}. 
Without loss of generality, we treat Alice's raw key ${\bm k}_{\rm A}  \in \{0,1\}^n$ as the reference. Bob's raw key ${\bm k}_{\rm B}$ is considered a noisy version of ${\bm k}_{\rm A}$, 
differing due to fabrication variability and environmental perturbations. 
An error-correcting code is employed to enable reconciliation by Bob. 

Alice applies a secure sketch (SS) function (based on a linear error-correcting code) to her key, yielding a syndrome ${\bm s}_{\rm A}$, often referred to as helper data \cite{Dod08}. 
SS ensures that ${\bm s}_{\rm A}$ reveals minimal information about ${\bm k}_{\rm A}$, while enabling correction of small errors. 
Alice sends the helper data ${\bm s}_{\rm A}$ to Bob over a public channel.

Let ${\cal C} \subseteq \{0,1\}^n$ be a linear $[n, k, d]$ binary code.  We may consider two different but equivalent constructions: \\
{\em 1. Code-Offset Construction.} 
Alice chooses a random  ${\bm x} \in \{0,1\}^k$, and computes the corresponding (random) codeword ${\bm c} := \mathcal{C}({\bm x})$. 
Subsequently, she computes the offset ${\bm s}_{\rm A}^{(1)} :=  {\bm k}_{\rm A} \oplus {\bm c}$, which is the SS for  ${\bm k}_{\rm A}$ and it is sent to Bob. 
Bob computes the vector ${\bm y} := {\bm k}_{\rm B} \oplus {\bm s}_{\rm A}^{(1)} = {\bm c} \oplus ({\bm k}_{\rm A}\oplus {\bm k}_{\rm B})$, and he decodes ${\bm y}$ to recover ${\bm c}$. 
If $\operatorname{dis}({\bm k}_{\rm A}, {\bm k}_{\rm B}) \le t$, with $t\leq \left\lfloor \frac{d - 1}{2} \right\rfloor$, then ${\bm k}_{\rm B} = {\bm k}_{\rm A} \oplus {\bm e}$ for an error vector ${\bm e}$ of weight at most $t$, and correct decoding will yield ${\bm c}$. Finally, Bob can compute ${\bm k}_{\rm A}$ as ${\bm c} \oplus {\bm s}_{\rm A}^{(1)}= {\bm k}_{\rm A} $.
\\
{\em 2. Syndrome Construction.} Let ${\bm H}$ be the parity-check matrix  of dimensions $ (n-k) \times n$. 
Alice computes the syndrome ${\bm s}_{\rm A}^{(2)} = {\bm k}_{\rm A}{\bm H}^{\rm T} = {\rm syn}({\bm k}_{\rm A})$, 
which is the SS for ${\bm k}_{\rm A}$, and it is sent to Bob. 
Bob computes the syndrome for his raw key ${\bm s}_{\rm B} ={\bm k}_{\rm B} {\bm H}^{\rm T}$, and subsequently he estimates the error 
syndrome  ${\bm s}_{\rm err} = {\bm s}_{\rm A}^{(2)} \oplus {\bm s}_{\rm B} =\left ( {\bm k}_{\rm A} \oplus {\bm k}_{\rm B} \right )  {\bm H}^{\rm T}= {\bm e} {\bm H}^{\rm T}$.
His task is to find unique ${\bm e} \in\{0,1\}^n$ of Hamming weight less than or equal to $t$, such that  ${\bm s}_{\rm err} = {\bm e}{\bm H}^{\rm T}$. 
Then he recovers ${\bm k}_{\rm A}$ as  ${\bm k}_{\rm A}= {\bm k}_{\rm B} \oplus{\bm e}$.
Note that if $\operatorname{dis}({\bm k}_{\rm A}, {\bm k}_{\rm B}) \le t$, then ${\bm e}$ has weight at most $t$, and its syndrome uniquely determines ${\bm e}$ under the decoding radius. 

The two constructions are equivalent. In construction 1, we have that the random offset ${\bm s}_{\rm A}^{(1)} $ is a member of the coset  ${\bm k}_{\rm A} \oplus {\cal C}$, and can be converted to the syndrome for ${\bm k}_{\rm A}$.  In particular, for a given  ${\bm s}_{\rm A}^{(1)} $, we have  ${\rm syn}( {\bm s}_{\rm A}^{(1)}) = {\bm s}_{\rm A}^{(1)}{\bm H}^{\rm T}  = ( {\bm k}_{\rm A} \oplus {\bm c}){\bm H}^{\rm T}  = {\bm k}_{\rm A} {\bm H}^{\rm T} =  {\rm syn}({\bm k}_{\rm A}) = {\bm s}_{\rm A}^{(2)} $. 
Moreover, given  ${\bm s}_{\rm A}^{(2)} $ in the second construction, one can sample a random representative of the coset ${\bm k}_{\rm A} \oplus {\cal C}$, 
which plays the role of the offset in construction 1. 
Indeed, we can choose at random a vector ${\bm v} \in \{0,1\}^n$ such that 
${\rm syn}( {\bm v})  = {\bm v} {\bm H}^{\rm T}=  {\bm s}_{\rm A}^{(2)}$. Given that ${\rm syn}( {\bm v} - {\bm  k}_{\rm A}) = 0$, this random vector will be of the form ${\bm k}_{\rm A} \oplus {\bm c}$, for some codeword ${\bm c}\in{\cal C}$.  
We see therefore that both constructions identify the same coset ${\bm k}_{\rm A} \oplus {\cal C}$:  the offset construction uses a representative of the coset for the identification, whereas the other construction uses the syndrome, which uniquely indexes the leader of the coset. 
 Since both identify the same coset, they support the same decoding logic and ensure recovery from any ${\bm k}_{\rm B}$ 
within Hamming distance $t$ from  ${\bm k}_{\rm A}$. 

By adopting either of the two constructions, Bob will recover Alice's key with high probability. Given the equivalence of the two 
constructions, from now on we assume that one of them is used in the protocol, and we denote by ${\bm s}_{\rm A}$  the helper data that Alice sends to Bob (the superscript is not necessary anymore).  
As a final step the two users have  to eliminate any information that may have leaked to an adversary, primarily through the transmission of  ${\bm s}_{\rm A}$.
To this end, Alice and Bob apply privacy amplification to their reconciled keys, obtaining a shared secret key with strong security guarantees \cite{PA95}. 
Typically, privacy amplification involves the application of hash functions by Alice and Bob. 

At this stage, we have presented a protocol for generating and distributing a common key between two honest users possessing twin PUFs. Before conducting a quantitative analysis of the protocol's main steps, we must define the adversarial model adopted throughout this work.

\subsection{Adversarial model}

We consider a passive adversary that has full access to the public classical communication exchanged during the protocol, but does not have physical access to the PUF devices held by the legitimate users. In particular, we assume that the adversary cannot query the PUFs, does not possess a physical clone, and has no access to side-channel information (e.g., optical emissions or hardware leakage) beyond the intended outputs. Under these assumptions, the adversary is limited to observing helper data and classical messages exchanged between Alice and Bob. While the main analysis assumes a passive adversary, we also briefly discuss the effect of active attacks on the classical communication, such as man-in-the-middle attack, for completeness. 

Machine-learning techniques constitute a plausible threat for many PUF-based cryptographic protocols. Typically such attacks require access to a large number of challenge–response pairs. In the present setting, the adversary cannot query the PUFs and does not observe the raw responses, but only publicly communicated syndromes. Under these assumptions, such attacks are not applicable. They may become relevant only in scenarios involving repeated protocol executions combined with significant information leakage, for example due to excessive syndrome exposure or additional side-channel information.

The security analysis presented in this work applies to a single execution of the protocol. In practice, successive executions should employ independent challenges, in which case the corresponding raw keys are expected to be statistically independent to a good approximation, and the information leakage associated with each execution is already accounted for by the reconciliation and privacy-amplification analysis. A more challenging scenario arises if identical or strongly correlated challenges are reused across multiple protocol executions. In that case, the publicly exchanged helper data may themselves become correlated and could potentially reveal additional information about the underlying raw keys. A rigorous treatment of such multi-session attacks would require an explicit statistical model of repeated protocol executions, and lies beyond the scope of the present work.

We note that side-channel attacks in optical systems may arise through unintended emissions, reflections, or hardware-level leakage. To the best of our knowledge, no generic model for such attacks exists in the context of optical PUFs; thus, mitigating their effects depends on the specific hardware implementation. It is worth emphasizing, however, that unlike QKD protocols, our scheme does not involve the exchange of optical signals between the users. Because all communication takes place over a classical channel and the optical PUFs operate within controlled, appropriately shielded environments, an adversary is denied unintended optical access, and cannot probe the devices via external challenges or remote measurements. This assumption of protected hardware is standard in physical security settings, and is generally less restrictive than those in many PUF-based entity-authentication schemes, where the device is intentionally exposed to external interrogation. Ultimately, this isolation of the physical PUFs from the communication channel significantly constrains external adversaries, allowing our analysis to focus on the information-theoretic aspects of key extraction.

Finally, in this work we assume a trusted and honest authority responsible for fabricating the PUF instances, which prepares exactly two correlated devices, referred to as twin PUFs. In principle, however, a manufacturer could produce more than two such correlated PUF instances, thereby enabling cryptographic protocols involving multiple users. At the same time, relaxing the two-instance assumption raises important security considerations for two-party cryptographic protocols (including the one presented here), as an adversary might attempt to obtain one of the additional instances and exploit it to compromise the protocol. This issue warrants careful examination, as it introduces new threat models beyond the standard two-party setting. The feasibility of producing multiple correlated PUF instances is expected to depend strongly on the underlying physical platform and fabrication process, and must therefore be evaluated on a case-by-case basis. Consequently, such an analysis lies beyond the scope of the present work.

Having specified the protocol for key agreement and the adversarial model, we proceeded to investigate the error rate in the raw keys of Alice and Bob for different levels of fabrication and environmental noise. 
This allowed us to evaluate the amount of redundancy required for error correction, and to assess the resulting length of the shared secret key after privacy amplification. 
For each level of fabrication noise, we generated many twin PUFs, and for each PUF we conducted repeated environmental trials (realizations). 
In order to quantify the response of the twins to the same challenge, we considered the bit error rate (BER), to be denoted as $Q$. This is an experimentally accessible quantity, and is computed as the normalized 
Hamming distance between the binary keys ${\bm k}_{\rm A}$ and ${\bm k}_{\rm B}$, which are generated from the speckles of the twins, using the Gabor filters and median thresholding.

\subsection{Error reconciliation}
\label{sec4a}

By deploying the twin PUFs Alice and Bob can essentially establish a noisy binary channel between them. 
Let $R := k/n$ be the rate of the error correcting code to be used for error reconciliation, 
where $k$ is the information bits (the length of the reconciled key in our case). 
According to Shannon's theory, a fundamental upper bound for the code rate is 
\be
R\leq 1- H_{\rm bin}(Q):=C
\label{shannon:eq}
\ee
where  $C$ is the channel capacity (in bits per channel use), and $H_{\rm bin}$ is the binary entropy. 

Shannon's limit gives the highest possible rate, achievable by any error-correcting code for a given BER, assuming ideal coding. 
It also specifies the smallest possible amount of helper data (also known as redundant or parity bits) required by the code, 
through the relation 
 \be
r := n-k = n\times (1-R)\geq n(1- C). 
\label{Lh:def}
\ee
However, practical codes do not achieve these bounds exactly. 
In general, it is desirable to keep redundancy low (i.e., to have $C$ large), because higher redundancy reduces the amount of key material available for privacy amplification.

\begin{figure*}[!t]
\centering
\includegraphics[width=6.in]{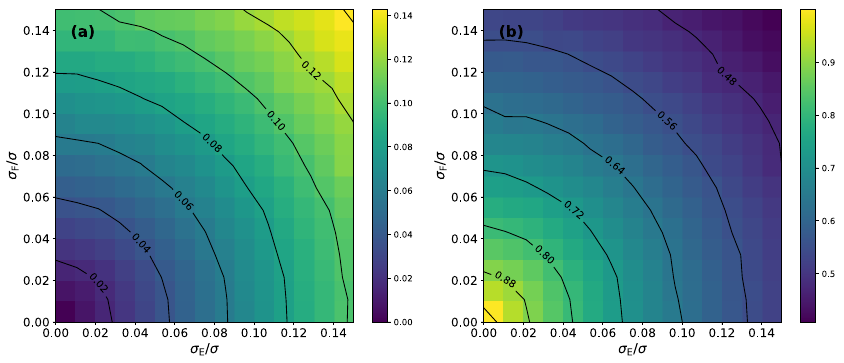}
\caption{The maximum BER $Q_{\rm \max}$ (a) and the corresponding channel capacity $C_{\rm \min}$ (b), are shown for different standard deviations of fabrication and environmental noise. The results have been obtained for 200 independent twins and 200 independent realizations for each twin. The fabrication and environmental noise have been treated as described in Sec. \ref{sec2} . Other parameters: ${\cal N} = 10^4$, $l/L = 0.2$, $1-$bit SLM, transmission geometry.}
\label{fig_2}
\end{figure*}

In order to correct $t$ errors in a raw key of length $n\gg1 $, we need an error-correcting code 
with minimum distance $d_{\min}\geq 2t+1$. Of course, the number of errors varies from realization to realization, 
due to fabrication and/or environmental noise. Throughout this work, we have adopted the worst-case 
scenario by considering  the maximum recorded BER $Q_{\rm max}$ in a large number of realizations, for different levels of noise $\sigma_{\rm F},\sigma_{\rm E}\in[0, 0.15]$. 
As shown in Fig. \ref{fig_2}(a), $Q_{\rm max}$ increases with increasing noise levels, but the growth is slow, and $Q_{\rm max}\lesssim 0.1$ for $\sigma_{\rm F},\sigma_{\rm E} \lesssim 0.14$.

By choosing a code that can correct $t\geq n\times Q_{\max} := t_{\max}$, we ensure that the probability of failure is negligible, because it is highly unlikely to obtain more than $ t_{\max}$ errors. 
Moreover, in plot \ref{fig_2}(b) we show the capacity $C_{\min} := 1- H_{\rm bin}(Q_{\max})$, 
which marks the highest rate that can be achieved by an error-correcting  
code that corrects $ t\geq t_{\max}$ errors, with vanishing probability of failure.  
Note that $C_{\min}$ decreases with 
increasing $\sigma_{\rm F}$ and $\sigma_{\rm E}$, because  $Q_{\max}$ increases.
However, for $\sigma_{\rm F},\sigma_{\rm E} \leq 0.14$ we find $C>0.56$ and according to Eq. (\ref{Lh:def}), the required redundancy is bounded from below by a quantity which does not exceed $0.44n$.  
Of course, this refers to ideal coding, and practical codes do not achieve the Shannon bound exactly. 

One important class of cyclic codes with  applications also in PUFs, is BCH codes \cite{Mes18,Sha12}.
For any positive integers $m\geq 3$ and $t<2^{m-1}$, there exists a binary BCH code with block length $n=2^m-1$, number of parity check bits $r=n-k\leq m t =t \log_2(n+1) $, and minimum distance $d_{\rm \min}\geq 2t+1$ (see chapter 6 in \cite{ECbook}).  Based on the numerical results depicted in Fig, \ref{fig_2}, we can choose for example $Q_{\max} \simeq 0.07$, which for $n=1023$ implies 
$t \lesssim n Q_{\max}  \simeq 71.61$. 
Hence, in this case, no more than $r=720$ bits of helper data would need to be transmitted (corresponding to $70\%$ redundancy), and a BCH  code with rate $ R \gtrsim 0.3$  would suffice.  BCH codes are not capacity-achieving and therefore operate below the Shannon limit for the corresponding noise levels. More advanced coding schemes, such as LDPC, polar or turbo codes \cite{ECbook2}, can approach channel capacity and would in principle allow operation at higher error rates or improved efficiency. However, BCH codes offer deterministic decoding, low implementation complexity, and are well suited to the moderate block lengths typical of PUF-based systems. Moreover, their fixed redundancy enables a transparent and explicit accounting of information leakage through the syndrome, which is essential for the information-theoretic security analysis carried out here. At higher noise levels (e.g., when $\sigma_{\rm F}, \sigma_{\rm E}\gtrsim 0.14$, corresponding to $Q_{\max} > 0.1$ in our simulations), achieving reliable reconciliation would require the use of near-capacity codes with substantially higher redundancy. The high-noise regime will be discussed in more detail in Sec. \ref{sec4c}. 

The above results have been confirmed for various combinations of parameters, and besides unimportant quantitative changes, the overall picture remains the same. 
That is, the error rate in the raw keys of Alice and Bob remains sufficiently low under moderate levels of fabrication and environmental noise 
$(\sigma_{\rm F}, \sigma_{\rm E}\lesssim 0.14)$, allowing for error reconciliation with relatively low redundancy overhead. Moreover, the larger ${\cal N}$ is, the more modes can be controlled, 
and this seems to improve the robustness of the system (lower error rate for given combination of noise variances). 
The combination of stronger environmental fluctuations and greater fabrication variability significantly increases the burden on error correction, thereby challenging the feasibility of efficient key generation without tighter 
process control or more advanced coding schemes.

\subsection{Privacy amplification and secret-key length}
\label{sec4b}

In the twin-PUF key agreement protocol, the length of the final secret key  is determined by the entropy of the raw (noisy) keys produced by the PUFs, 
as well as by the information leakage resulting from the public transmission of helper data (syndrome) during error reconciliation. 

After successful error reconciliation, Alice and Bob apply \emph{privacy amplification} \cite{PA95} to eliminate any residual information potentially available to an adversary. According to the \emph{leftover hash lemma}, when a two-universal hash function is used, the length $\ell$ of the final secret key satisfies \cite{KeyRate2,KeyRate3}
\begin{equation}
\ell \ge H_{\min}(\bm{k}_{\rm A} \mid \bm{s}_{\rm A}) - 2\log\!\left(\frac{1}{\varepsilon}\right),
\end{equation}
where $H_{\min}(\bm{k}_{\rm A} \mid \bm{s}_{\rm A})$ denotes the conditional min-entropy of Alice’s raw key given the publicly available helper data (syndrome) $\bm{s}_{\rm A}$, and $\varepsilon \ll 1$ is the security parameter.

Let $p_{\rm guess} (\bm{k}_{\rm A} \mid \bm{s}_{\rm A}) $ be the maximum (worst case) probability for a third party to guess $\bm{k}_{\rm A}$, using the syndrome $ \bm{s}_{\rm A}$ and the optimal strategy.  Then we have \cite{KeyRate2}
\bea
p_{\rm guess}(\bm{k}_{\rm A} \mid \bm{s}_{\rm A}) = 2^{-H_{\min}(\bm{k}_{\rm A} \mid \bm{s}_{\rm A}) },
\eea
which allows us to write 
\bea
H_{\min}(\bm{k}_{\rm A} \mid \bm{s}_{\rm A}) = -\log_2[p_{\rm guess}(\bm{k}_A \mid \bm{s}_{\rm A})].
\eea

The public transmission of the syndrome $\bm{s}_{\rm A} \in \{0,1\}^r$ leaks information about the raw key. In general, this leakage can be bounded using the standard min-entropy inequality
\begin{equation}
H_{\min}(\bm{k}_{\rm A} \mid \bm{s}_{\rm A}) \ge H_{\min}(\bm{k}_{\rm A}) - \log |\mathcal{S}|,
\end{equation}
where $\mathcal{S}$ is the set of possible values for  $\bm{s}_{\rm A}$ with  $|\mathcal{S}| \leq 2^r$, and  equality holds only 
in the case of a full $r$-bit syndrome. Hence,
\begin{equation}
H_{\min}(\bm{k}_{\rm A} \mid \bm{s}_{\rm A}) \ge H_{\min}(\bm{k}_{\rm A}) - r.
\end{equation}

Besides the syndrome, in the protocol under consideration the adversary has access only to the challenge exchanged between Alice and Bob. 
Assuming that (i) the adversary does not have access to a copy of either of the twin PUFs;
(ii)  the raw keys generated by the PUFs are uniformly distributed over $\{0,1\}^n$; and (iii)
the challenge does not convey any information about the raw keys, the best an adversary can do  is to guess $\bm{k}_{\rm A}$, 
using  ${\bm s}_{\rm A}$ and her knowledge on the details of the error-correcting code used by Alice and Bob. 
Under the additional assumption that (iv) the syndrome mapping  (i.e., $\bm{k}_{\rm A} \mapsto \bm{s}_{\rm A}$)  is linear, so that each syndrome corresponds to a coset of size $2^{n-r}$ and all cosets are equally likely, the conditional distribution of $\bm{k}_{\rm A}$ given $\bm{s}_{\rm A}$ is uniform over the corresponding coset. In this ideal case,
\begin{equation}
p_{\mathrm{guess}}(\bm{k}_{\rm A} \mid \bm{s}_{\rm A}) = \frac{1}{2^{n-r}},
\end{equation}
and therefore
\begin{equation}
H_{\min}(\bm{k}_{\rm A} \mid \bm{s}_{\rm A}) = n - r.
\end{equation}
Substituting into the leftover hash lemma yields
\bea
\ell \geq n-r- 2\log\left(\frac{1}{\varepsilon}\right).
\label{key_length:eq}
\eea
We see therefore that under these assumptions, in each execution of the protocol (i.e., for each challenge), the publicly transmitted syndrome reveals at most $r$ bits of information about the raw key, where $r$ is the number of parity bits of the error-correcting code. This follows from the fact that the syndrome takes values in a set of size at most $2^r$, implying that it can reduce the min-entropy of the raw key by at most $r$ bits.
Hence, the final key length is constrained by the redundancy $r$ of the error-correcting code and the desired security level $\varepsilon$, with 
the code rate $R$ playing a central role in this tradeoff, as dictated by inequality (\ref{Lh:def}). In the example with the BCH code mentioned above, $n=1023$, $r = 720$, and for $\varepsilon = 2^{-25}$ we have $\ell\geq 253$ bits. 

More generally, relaxing the assumption of uniform keys, the extractable key length satisfies
\begin{equation}
\ell \ge H_{\min}(\bm{k}_{\rm A}) - \mathrm{leak} - 2\log\!\left(\frac{1}{\varepsilon}\right).
\end{equation}
Here, $\mathrm{leak}$ denotes the total information about the raw key revealed to the adversary under the assumed attack model. For the adversarial model under consideration, and assuming that the challenge is uncorrelated with the raw key, this leakage is bounded from above by the total number of bits publicly revealed during the error-reconciliation phase. For linear codes (e.g., BCH or LDPC), this corresponds to the number of parity bits, $r=n-k$. For more general reconciliation protocols, $r$ represents the total length of all publicly exchanged information and may exceed $n-k$.
In practice, the min-entropy $H_{\min}(\bm{k}_{\rm A})$ 
depends strongly on the details of the particular implementation under consideration, and 
must be estimated empirically to account for bias and correlations in the generated binary keys.  
In particular, it can be estimated using standard widely used 
cryptographic tools for the evaluation of random-number generators  (e.g., NIST Statistical Test Suite).  
Using this suite of tests, the key generated from a given PUF is being benchmarked against the statistical behavior of a true random number 
source. Akriotou and co-workers \cite{Akr18} have reported on an  optical PUF, which can generate binary keys with min entropy that exceeds 0.92,  
which justifies our assumption of nearly uniform ${\bm k}_{\rm A}$. 

The above analysis assumes that privacy amplification is performed using a two-universal hash function, as required by the leftover hash lemma, which provides 
an \emph{information-theoretic} guarantee on the secrecy of the final key. Other, fixed cryptographic hash functions such as SHA-256 can be also used  
for privacy amplification. These are not formally two-universal, and their security is not covered by the same information-theoretic guarantees. Instead, their use relies 
on \emph{computational assumptions} and the belief that they behave like pseudorandom functions or random oracles when keyed with high-entropy inputs. 
As a result, the leftover hash lemma no longer applies directly in this case, and the corresponding security guarantees shift from provable statistical bounds to empirical and  cryptographic robustness. 

\subsection{High-noise regime and reconciliation limits}
\label{sec4c}

It is also instructive to consider the key length that is predicted by the Shannon limit for a given BER Q. 
Assuming ideal capacity-achieving reconciliation, one would have $r_{\min} =  n H_{\rm bin}(Q)$ redundant bits, 
which, upon substitution into Eq.~(\ref{key_length:eq}), leads to the optimistic estimate 
\begin{equation}
\ell_{\rm ideal}
\simeq n [1-H_{\rm bin} (Q)]+2\log(\varepsilon). 
\label{key_length_ideal:eq}
\end{equation}
Using the standard BCH redundancy bound and choosing $t\simeq nQ$, Eq. (\ref{key_length:eq}) gives the conservative estimate 
\be
\ell_{\rm BCH}\geq  n[1-Q\log_2(n+1)]+2\log_2(\varepsilon),
\label{key_length_bch:eq}
\ee
which is positive for 
\be
Q\leq \frac{n-2|\log_{2}(\varepsilon)|}{n\log_2(n+1)}.
\label{Qbound_bch:eq}
\ee
For the illustrative parameters $n=1023$ and $\varepsilon=2^{-25}$, this estimate yields $Q\lesssim 0.095$. 
The BCH construction considered in Sec. \ref{sec4a} is therefore expected to be effective only in the low-BER regime
 (i.e., $Q\lesssim 0.1$), and cannot efficiently support the much larger BER values reported in Ref. \cite{Mar22}. 
Operation in such a high-noise regime would require reconciliation protocols that operate significantly closer to the Shannon limit, such as LDPC or polar codes. For such near-capacity reconciliation schemes, one may write
\[
r \simeq fnH_{\rm bin}(Q),
\]
where \(f\ge 1\) denotes the reconciliation inefficiency (Shannon limit is achieved for $f=1$). Thus Eq. (\ref{key_length:eq}) reads 
\be
\ell\geq  n[1-fH_2(Q)]+2\log_2(\varepsilon).
\label{key_length_ldpc:eq}
\ee

Table~\ref{tab:key_length_vs_ber} illustrates the impact of the BER on
 the achievable secret-key length.
The BCH construction considered in our example above  is effective primarily in
the low-BER regime, while its redundancy rapidly becomes prohibitive as
the BER increases, thereby leading to negative key rates. By contrast, reconciliation schemes operating close
to the Shannon limit, such as suitably designed LDPC or polar codes,
can still yield positive secret-key lengths at substantially larger BER
values. At $Q=0.3$, positive asymptotic key generation requires
$f<1/H_{\rm bin}(0.3)\approx1.135$. For finite block length, however,
the penalty due to privacy amplification  must also be taken into account. Thus for
$n=1023$ and $\varepsilon=2^{-25}$, the corresponding condition becomes
$f < 1.079$. This explains why the column corresponding to $f=1.05$ in Table~\ref{tab:key_length_vs_ber} still yields a positive key length at $Q=0.3$, whereas the column corresponding to $f=1.10$ does not.

Finally, it should pointed out that the dependence on the raw-key length differs considerably for BCH and near-capacity reconciliation schemes. 
For the BCH construction considered here, Eq. (\ref{Qbound_bch:eq}) reduces to  $Q <[\log_2(n+1)]^{-1}$ for $n\gg|\log_2\varepsilon|$, 
indicating that increasing the raw-key length alone does not significantly improve the tolerable BER. 
By contrast, for LDPC or polar codes the asymptotic condition $f<1/H_{\rm bin}(Q)$ is independent of $n$, while finite-size corrections scale as $1/n$. 
Consequently, larger raw-key lengths can substantially improve the performance of near-capacity reconciliation schemes in the high-BER regime. 
For example, as shown in Table~\ref{tab:key_length_vs_ber} for $Q=0.3$ and $f=1.1$, no positive key is expected for $n=1023$. 
However, Eq.~(\ref{key_length_ldpc:eq}) yields $\ell\approx 13$ bits when the raw-key length is doubled to $n=2046$. 
This illustrates that increasing the raw-key length can compensate for the finite privacy-amplification penalty in near-capacity reconciliation schemes, whereas the BCH construction considered above remains fundamentally restricted to the low-BER regime.

\begin{table}[t]
\centering
\caption{Illustrative secret-key length estimates for $n=1023$ and
$\varepsilon=2^{-25}$. For the BCH estimate we use Eq. (\ref{key_length_bch:eq}). 
For near-capacity reconciliation (LDPC/polar codes) we use Eq. (\ref{key_length_ldpc:eq}) with representative inefficiencies $f=1.05$ and $f=1.10$. 
The last two columns show the corresponding Shannon-limit redundancy $r_{\rm min}= nH_{\rm bin}(Q)$ and the optimistic secret-key benchmark $\ell_{\rm ideal}=n-r_{\rm min}+\log(\varepsilon)$. 
Negative values indicate that no positive key is obtained under the corresponding assumptions. 
}
\label{tab:key_length_vs_ber}
\begin{tabular}{c | c | c c | c}
\hline
BER $Q$ & $\ell_{\rm BCH}$ & $\ell_{f=1.05}$ & $\ell_{f=1.10}$  & $\ell_{\rm ideal}$
\\
\hline
0.01  & 863 & 886 & 882  & 891
\\
0.03  & 663 & 764 & 754  & 773
\\
0.05  & 453 & 665 & 651  & 691
\\
0.07 & 253 & 580 & 561  & 608
\\
0.10  & - & 469 & 445  & 494
\\
0.15  & - & 318 & 287  & 343
\\
0.20  & - & 198 & 161  & 223
\\
0.30  & - & 27 & - &  72
\\
\hline
\end{tabular}
\end{table}

\subsection{The role of authentication}
\label{sec4d}

In the twin-PUF key distribution protocol under consideration, classical communication between the users occurs  in at most three different stages of  the protocol. 
Initially, Alice sends Bob a randomly chosen challenge, and later she also sends the helper data (syndrome) in order for Bob to be able to reconstruct her key.  
Finally, during the privacy amplification stage, communication may again be required, specifically when \emph{keyed} hash functions are used and the users 
must agree on a randomly selected  function instance.  
In this case, one of the users, say Alice, chooses at random a  function from the two-universal hash family, and informs Bob about her choice.  

If communication takes place over an \emph{unauthenticated channel}, an active adversary may tamper with the messages, thereby compromising the protocol. 
For instance, altering the challenge or the syndrome sent to Bob may desynchronize the twin PUFs, preventing the establishment of a common key between the users. Even more critically, 
if the adversary can influence the hash function selection during privacy amplification, he/she could bias e.g., Bob’s output key (potentially learning it), but without ever knowing Alice’s key. 
This breaks the confidentiality of the protocol, even though the underlying physical sources remain secure.
By contrast, if a \emph{keyless} extractor or a fixed cryptographic hash function (e.g., SHA) is used for privacy amplification, no communication is needed at this stage, 
and the associated attack  is eliminated. However, as discussed above, this  comes at the cost of losing strong information-theoretic guarantees, 
since the leftover hash lemma applies strictly to two-universal hash families.

These considerations highlight the importance of authenticated communication in any key agreement protocol, including the ones involving PUFs,  where even if the 
randomness source is considered tamper-resistant,  the communication channel may not be.

\subsection{Seeding  QKD systems with twin-PUF generated key}
\label{sec5}

QKD enables two parties to generate a shared secret truly random key with information-theoretic security, even in the presence of an unbounded adversary. 
However, the classical communication channel used during QKD must be \emph{authenticated}; otherwise, a man-in-the-middle attack can compromise the entire protocol. 
To this end, any QKD system requires that the two honest users share a common secret key \emph{before} running the protocol for the first time. 
The secure distribution of this initial pre-shared key is therefore a critical issue for the overall security of QKD. 
Proposed solutions include storing a pre-shared key within the QKD hardware, using post-quantum cryptographic (PQC) schemes \cite{QKD-PQC}, or leveraging PUFs \cite{NikFis24,Konteli26}.

The first option (i.e., pre-storing a key in the QKD devices or other portable storage devices) is problematic, as it provides no mechanism for key refresh and lacks formal security guarantees. Hybridization of QKD with PQC offers an efficient approach to key bootstrapping and management, but reduces the level of security from information-theoretic to computational. In such cases, the distributed final key is secure only as long as the underlying PQC scheme remains unbroken during the execution of the QKD  protocol (a condition that can be guaranteed only under computational assumptions). Importantly, however, if the adversary is not present during the execution of the hybrid protocol, he/she cannot later obtain the final key, even with unlimited resources.

The use of PUFs to generate and distribute pre-shared keys has emerged as an alternative lightweight, hardware-rooted solution \cite{NikFis24,Brz11}. The core idea is to associate each QKD device with a unique disordered token and a PUF \cite{NikFis24}. Because the PUFs are independent, their responses to the same challenge are uncorrelated. To enable key agreement, a trusted authority (e.g., the QKD manufacturer) stores in a secure database, the XOR value of the PUF responses to selected challenges. 
These stored values are later retrieved by the  owners of the QKD devices, in order to establish a common pre-shared key for initializing the QKD protocol. 
The security of the pre-shared key—and hence of the QKD system—relies on two key assumptions: (i) the numerical keys produced by the PUFs are indistinguishable from truly random binary strings, and (ii) the PUF tokens remain physically secure and  inaccessible to adversaries.
The necessity for a database in the scheme of Ref. \cite{NikFis24,Konteli26}, stems from the independence of the PUFs, which prevents direct correlation between their responses to the same challenge. This requirement can be relaxed by using twin PUFs, thereby enabling synchronized derivation of a pre-shared key via the protocol discussed in Sec.~\ref{sec3}. In this case, it is important to  employ two-universal hash functions during the privacy amplification stage.  The associated classical communication during the establishment of the pre-shared key need not be authenticated.
This is because, as discussed above, an adversary who tampers with the classical communication may cause Alice and Bob to derive mismatched pre-shared keys. 
However, such a mismatch will almost certainly cause the QKD protocol to abort, thereby informing the users of the failure of the twin-PUF key distribution at the outset. 
Crucially, the security of the QKD session remains uncompromised, as no fresh key has been generated and no key material has been leaked to  adversaries.
We emphasize that this argument does not extend to general use of our protocol. Outside the QKD-seeding scenario, authenticated classical communication remains necessary to ensure security of the protocol against active adversaries.

\section{Discussion and Conclusion}

We have discussed the generation and distribution of secure keys between two honest users by leveraging twin optical PUFs. 
Taking into account fabrication imperfections and environmental fluctuations, we identified the level of noise that allows for the establishment of  a secret key. 
The protocol does not require transmission and detection of photonic states, and thus it is not limited by the distances between the two users. Moreover, in contrast 
to previous work in the field \cite{Horstmayer13,NikFis24,Konteli26}, there is no need for storing the PUF responses to selected challenges in a secure database. 

To derive a tractable and meaningful lower bound on the final secret-key length, we adopted a set of standard assumptions commonly used in PUF applications. 
First, we assumed that the challenge does not convey any information about the output of the PUFs which is uniformly distributed over $\{0,1\}^n$, reflecting the ideal behavior of strong PUFs.  Second, we assumed that the helper data (syndrome) are generated from each PUF output using a linear code.
Third, we assumed that the adversary has access only to the publicly revealed helper data (syndrome) and does not possess any side-channel information or access to either of the PUF instances. Finally we assumed the twin PUFs are fabricated by a trusted honest authority. 
While these assumptions provide a clean analytical and by no means unrealistic framework, relaxing them would require a more detailed and potentially system-specific modeling of the PUF behavior, the error-correction, and the adversary’s capabilities, in order to obtain a 
valid lower bound on the conditional min-entropy and, consequently, on the final key length.  

Our analysis of error reconciliation and the final secret key length focuses on a single execution of the protocol. However, by running the protocol multiple times using independently chosen random challenges (and assuming that the above mentioned assumptions about the PUFs hold) one can concatenate the resulting secret keys to obtain a longer key with the same level of security. Provided that independently chosen challenges generate statistically independent raw responses, each execution contributes fresh entropy, and privacy amplification ensures that the leakage associated with each execution remains negligible.

In our protocol, neither the raw key nor the final secret key is ever revealed. The only information available to an adversary arises from the publicly transmitted syndrome. As long as this leakage is sufficiently small, it can be effectively removed through privacy amplification. 
The protocol can be executed multiple times using fresh challenges, resulting in independent raw responses for each run. 
Under this condition, information leakage from helper data does not accumulate in a way that compromises security, as it is removed through privacy amplification. 
However, repeated use of the same challenge leads to multiple observations of helper data associated with the same raw response, allowing an adversary to accumulate information about that response. 
In principle, information accumulated across multiple protocol executions could be exploited to approximate the behavior of the PUFs, for example using machine-learning techniques. Such attacks typically rely on access to a large number of challenge–response pairs in order to learn the underlying mapping between challenges and responses. In the present setting, the adversary cannot query the PUFs and does not observe the raw responses, but only publicly communicated helper data. 
Under these conditions, such attacks have not been shown to be effective within the adversarial model considered here, and may become relevant only in scenarios involving repeated protocol executions with additional information leakage, which are beyond the scope of this work. 
Therefore, as long as each challenge–response pair is used only once and the chosen secure sketch ensures sufficiently small information leakage, the protocol remains secure under the assumptions of our model.

In its present form, the proposed protocol is not an alternative to QKD, as the security model of synchronized optical PUFs differs fundamentally from that of QKD. QKD protocols rely on the principles of quantum mechanics to guarantee information-theoretic security against eavesdropping, albeit with the caveat that they require specialized quantum channels and are subject to loss, noise, and distance limitations. In contrast, the security of the proposed PUF-based protocol is rooted in the unclonability and unpredictability of disordered optical structures, and it does not require the transmission of quantum states. This makes PUF-based key agreement potentially more practical and cost-effective in scenarios where QKD deployment is technologically or economically prohibitive. 

To the best of our knowledge, this is the first work to combine a twin optical-PUF physical model with an explicit information-theoretic key-agreement framework incorporating secure sketches, error reconciliation, privacy amplification, and quantitative leakage analysis.
 We further quantify information leakage and secret-key rates under realistic assumptions using a general platform-independent stochastic noise model.
 The proposed protocol is amenable to experimental verification using existing twin optical-PUF technologies. 
The analysis presented here should be interpreted as identifying the noise regimes under which information-theoretically secure key agreement with twin PUFs becomes feasible. 
Hence, the present work complements existing experimental demonstrations of twin PUFs by providing a quantitative framework for assessing the impact of noise, reconciliation overhead, helper-data leakage, and privacy amplification on the achievable secret-key length.  In this sense, the derived BER thresholds are best viewed as performance targets for practical implementations rather than as limitations of the protocol itself. While some of the twin-PUF platforms reported in the literature currently operate at comparatively high BER values \cite{Mar22}, there is no theoretical reason to expect these values to be universal across all twin-PUF architectures or fabrication technologies. Improvements in device fabrication, interrogation schemes, feature extraction, and error-correction techniques can all contribute to reducing the effective BER, thereby  reducing the reconciliation overhead and increasing the achievable secret-key length. 
Another promising direction for future research is the combination of the twin-PUF paradigm with recently proposed reconfigurable optical PUF architectures \cite{Kwak2026}. Reconfigurable PUFs can generate multiple distinct optical responses from the same physical device through controlled modifications of the underlying scattering structure.
If such functionality could be realized in a twin-PUF setting, it would provide a mechanism for generating multiple independent shared randomness resources from the same pair of devices, thereby supporting repeated key-generation sessions without exhausting the available challenge space. 
The development of reconfigurable twin optical PUFs and the corresponding information-theoretic analysis constitute interesting topics for future investigation.

Although the physical model developed in this work is based on light scattering in correlated optical random media, the information-theoretic key-agreement framework is considerably more general. Once two physically correlated, not necessarily optical, PUFs generate raw binary keys characterized by a BER $Q$, the reconciliation and privacy-amplification procedures depend primarily on this quantity rather than on the underlying physical realization.  A representative example is provided by the CNT-based twin PUFs reported in Ref. \cite{Zhong22}. They exhibit a consistency of approximately 95\% between twin responses, corresponding to an estimated disagreement probability of approximately  $Q\approx0.05$.  Although the underlying physical platform differs substantially from the optical scattering systems considered in the present work, our key-exchange protocol can be also applied in the setup of Ref. \cite{Zhong22}. 
The low disagreement probability reported in Ref. \cite{Zhong22} suggests that efficient reconciliation can be achieved without resorting to near-capacity coding schemes. Moreover, this result illustrates that low disagreement probabilities can be achieved in twin-PUF architectures, and reinforces the view that the BER values reported in Ref.~\cite{Mar22} should not be regarded as fundamental limitations of the twin-PUF concept.

\ack{The publication of the article in OA mode was financially supported by HEAL-Link.}

\funding{This research was co-funded by the European Union under the Digital Europe Program grant agreement number 101091504.}

\section*{Conflicts of Interest}
The authors declare no conflicts of interest.

\data{All data that support the findings of this study are included within the article (and any supplementary files).}




\end{document}